\documentclass[twocolumn,prl,showpacs]{revtex4}
\usepackage{graphicx}
\usepackage{bm}
\usepackage{amsmath}
\usepackage{amssymb}
\usepackage{aas_macros}
\usepackage{subfigure}
\usepackage{hyperref}

\begin{document}

\title{A new Tolman test of a cosmic distance duality relation at 21 cm}
\author{Satej Khedekar}
\email{satejk@tifr.res.in}
\author{Sayan Chakraborti}
\email{sayan@tifr.res.in}
\affiliation{Tata Institute of Fundamental Research, 1 Homi Bhabha Road, Colaba, Mumbai - 400005, India.}

\begin{abstract}
Under certain general conditions in an expanding universe,
the luminosity distance ($d_L$) and angular diameter distance
($d_A$) are connected by the Etherington relation as $d_L=d_A (1+z)^2$. 
The Tolman test suggests the use of objects of known surface brightness, to test
this relation. In this letter, we propose the use of redshifted 21 cm signal from disk galaxies, 
where neutral hydrogen (HI) masses are seen to be almost linearly correlated
with surface area, to conduct a new Tolman test. We construct simulated catalogs
 of galaxies, with the observed size-luminosity relation and realistic redshift
 evolution of HI mass functions, likely to be detected with the planned Square
 Kilometer Array (SKA). We demonstrate that these observations may soon provide the best 
 implementation of the Tolman test to detect any violation of the cosmic distance duality relation. 
\end{abstract}


\maketitle

{\it Introduction}.---Observed fluxes and apparent sizes of cosmologically distant objects determine
their $d_L$ and $d_A$ respectively, which are tied by the \citet{1933PMag...15..761E} relation as
$d_L=d_A (1+z)^{2 \iffalse +\epsilon \fi}$. 
This relation holds true for photons traveling on unique null geodesics in an expanding universe
 described by a metric theory of gravity and is 
guaranteed by phase space conservation of photons along with Lorentz invariance.
Testing the Etherington relation is important because any violation of it
would be a {\it smoking gun for new physics}. At the same time, it is important
to distinguish between violations occurring due to new physics and
astrophysical effects; often the signatures of these effects may not be
distinguishable. Among the important astrophysical effects are dimming due to
the intergalactic dust and gravitational lensing effects \cite{2004PhRvD..69j1305B};
 while signatures of new physics could arise for example from interactions of 
photons with the dark sector.

\citet{1930PNAS...16..511T} suggested that measurements of angular extension,
luminosity and redshift of objects with a standard surface brightness may be
used to test this {\it cosmic distance duality} relation (CDDR) \footnote{The Etherington
 relation is purely a statement about differential geometry, while the distance duality relation
 has the additional requirements of photon conservation as well as the photons traveling
 on null geodesics \cite{2004PhRvD..69j1305B}.}.
In this letter, we propose 21 cm observations of disk galaxies as a new Tolman test
that promises a systematics-free and clean probe of the CDDR.
The advantage of our test over many of the existing ones is that it is immune to
many astrophysical uncertainties. We show that future observations
 like those with the planned SKA would easily be able to probe this relation up 
to $z \sim 1$. We also discuss possible strategies to obtain significantly
 tighter constraints on any possible violation of this relation.

{\it Existing Tests}.---\citet{2001AJ....121.2271S} performed optical photometry of nearby galaxies
and compared them with a sample of early-type
galaxies at high redshift \citep{2001AJ....122.1084L}, performing the Tolman surface brightness test.
The surface brightness was seen to be falling at a rate significantly slower
than the $\propto(1+z)^{-4}$ expected from the Etherington relation.
However, the authors suggested that this discrepancy may be 
reconciled with an expanding universe by invoking luminosity
 evolution of early-type galaxies.

Another way to test the CDDR is to make independent measurements of
the cosmic distances $d_L$ and $d_A$ using different objects. 
References \cite{2004PhRvD..69j1305B, 2009ApJ...696.1727M, 2010JCAP...10..024A}
provide upper limits on the {\it cosmic transparency} 
by observing the flux dimming of {\it standard candles}
\cite{2010ApJ...716..712A}
along with independent measurements of a cosmic distance scale through baryon acoustic oscillation
measurements \cite{2010MNRAS.401.2148P} and $H(z)$ \cite{2010JCAP...02..008S}
data. However, all these methods suffer from uncertainties due to absorption of optical 
photons by dust; and also the possibility of being affected by lensing magnification effects.

\citet{2011A&A...528L..14H} have also examined the consistency of the
CDDR through combined X-ray and SZ observations of galaxy clusters.
Although this method has the advantage of being unaffected by lensing magnification bias,
 their results depend largely on the cluster model and could suffer from unknown
systematics.

\begin{figure*}[ht]
  \centering
  \begin{minipage}[c]{0.2\textwidth}
    \hspace*{0.5cm}
    \includegraphics[width=\textwidth]{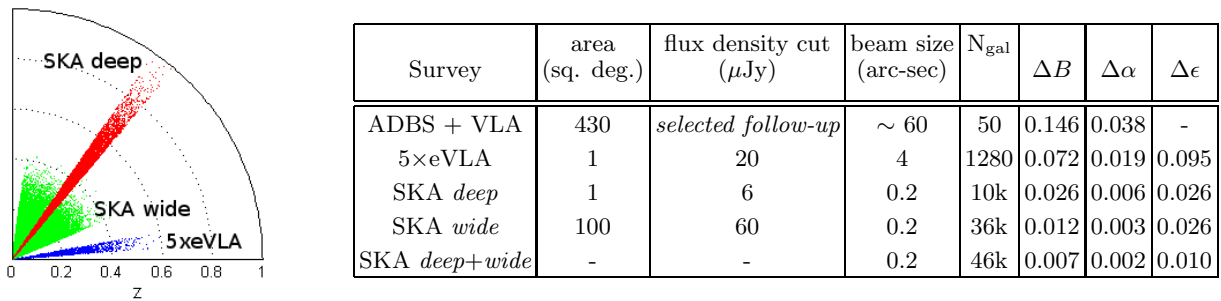}
  \end{minipage}
  \begin{minipage}[c]{0.75\textwidth}
    \hspace*{1cm}
    \begin{tabular}{|c|c|c|c|c|c|c|c|}
	\hline 
		& area & flux density cut & beam size & ${\rm N_{gal}}$ & & & \\  
	\raisebox{1.5ex}{Survey} & \raisebox{1.5ex}{(sq. deg.)} & \raisebox{1.5ex}{($\mu$Jy)} & \raisebox{1.5ex}{(arc-sec)} & & \raisebox{1.5ex}{$\Delta B$} & \raisebox{1.5ex}{$\Delta \alpha$} & \raisebox{1.5ex}{$\Delta \epsilon$} \\
	\hline
	\hline  
	ADBS + VLA  		& 430  	& {\it selected follow-up}	& $\sim$ 60 	& 50	& 0.146	& 0.038 & -  	\\
	5$\times$eVLA 		& 1	& 20				& 4		& 1280	& 0.072 & 0.019 & 0.095 \\
	SKA {\it deep} 		& 1 	& 6 				& 0.2		& 10k	& 0.026	& 0.006	& 0.026 \\
	SKA {\it wide} 		& 100 	& 60				& 0.2 		& 36k	& 0.012	& 0.003	& 0.026 \\
	SKA {\it deep{\rm +}wide}	& -	& - 				& 0.2		& 46k	& 0.007	& 0.002 & 0.010 \\
	\hline
    \end{tabular}
  \end{minipage}  
  \caption{The figure on the left represents the expected redshift distribution 
from some upcoming surveys. All angles in this figure are scaled up by a factor of 
4 for representative purposes. The survey specifications and the 1$\sigma$
marginalized constraints on the parameters $B$, $\alpha$ and $\epsilon$ are listed
in the table on the right.}
  \label{tab:both}
\end{figure*}

{\it Our proposed test at 21 cm}.---The Arecibo Dual-Beam Survey (ADBS) covered $\sim$ 430 deg$^2$ of the sky using the
Arecibo main beam with a velocity coverage of $-$654 to 7977 km s$^{-1}$ 
\citep{2000ApJS..130..177R} to detect the emission at 21 cm from galaxies. 
Out of the 265 identified galaxies, only
a few were resolved in the Arecibo observations.
\citet{2003ApJ...585..256R} made spatially resolved D-array
follow-up observations with the Very Large Array (VLA) for
84 of these galaxies. Of these, 50 well resolved sources 
were used \cite{2000ApJS..130..177R}
to determine their HI masses and surface areas. In this study
we use this sample to calibrate the size luminosity correlations
to be used in the generation of our mock catalogs.

HI masses and physical sizes of ADBS \citep{2000ApJS..130..177R} galaxies are found to be
consistent with a nearly constant average HI surface density
of the order $\sim10^7 \ {\rm M}_\odot$ kpc$^{-2}$ \citep{2003ApJ...585..256R}.
The HI masses of individual ADBS galaxies, spanning more than three orders of
magnitude, deviate by only $\sim0.13$ dex ($1\sigma$) from those
expected from a constant HI surface density. This surprising
relation has recently been explained by \citet{2011arXiv1103.2763C}
as the result of self-regulated star formation, driven by the
competition between gravitational instabilities in a rotationally supported disk and
mechanical feedback from supernovae. \citet{1997ApJ...481..703S} shows that supernovae
drive up the porosity of the gas disk and stabilize it by increasing the gas dispersion
velocity. Analytic \citep{1997ApJ...481..703S} and simulation 
\citep{2009ApJ...704..137J} results suggest that the gas
 dispersion velocity driven by supernova feedback is insensitive to the
 star formation rate (SFR) \footnote{The gas dispersion velocity changes by
 less than a factor of 2 for a 512 fold change in SFR \citep{2009ApJ...704..137J}.} as well as metallicity
 \footnote{According to Eq 5 \& 8 of \citep{2011arXiv1103.2763C} a factor 3 change in metallicity only gives
a 1\% change in the HI surface density.}. We find the redshift dependent
 effects in the evolution of the luminosity-area relation to be much lower
 than the intrinsic scatter in the HI masses.

The present galaxy sample lies on the fundamental line
\citep{2011arXiv1103.2763C} even though the sample probes a large 
range in size and metallicity from dwarf galaxies to giant spirals.
These objects with nearly constant
surface brightness which is both documented \cite{2003ApJ...585..256R} 
and understood \citep{2011arXiv1103.2763C}, then
provide us with astrophysical sources to give a direct measurement of 
the distance ratio $d_L/d_A$ at various redshifts.

The HI emission starts out at 21 cm radio in the rest frame of the
source. It gets redshifted out of the resonance line to longer wavelengths as the universe expands
during its travel to the observer. The universe is essentially transparent
at these wavelengths as their propagation is not affected by gas and dust.
Only HI can reabsorb the 21 cm emission before it is redshifted out of the
resonance line. But the absence of the \citet{1965ApJ...142.1633G} absorption
feature for quasars below $z\lesssim6$, confirm that there is hardly any
HI ($n_{\rm HI}$ $\lesssim 6 \times 10^{-11}$ atoms cm$^{-3}$)
along these lines of sight. Considering the cross-section of the 
21 cm hyperfine transition, we get the optical depth as
$\sim 10^{-7}$, which is negligible when compared to the intrinsic scatter. This is a clear
 advantage over other tests of the CDDR at visible wavelengths where the effect of dimming due
 to dust is inseparable from effects due to new physics.
Additionally, the use of the same objects to measure the ratio $d_L/d_A$ also 
means that there is no risk of lensing magnification bias, since the surface brightness 
of a single source remains preserved after gravitational lensing \cite{1964MNRAS.128..295R}.

{\it Simulated Galaxy Catalogs}.---To asses the potential of future radio observations in probing the 
CDDR using our proposed Tolman test, we create 
mock galaxy catalogs with a realistic redshift distribution that 
would be expected to be seen in these surveys (see Fig. \ref{tab:both}). 
We briefly outline the procedure involved in doing so.

\begin{figure*}[ht]
\centering
\subfigure[]{
  \includegraphics[width=8cm]{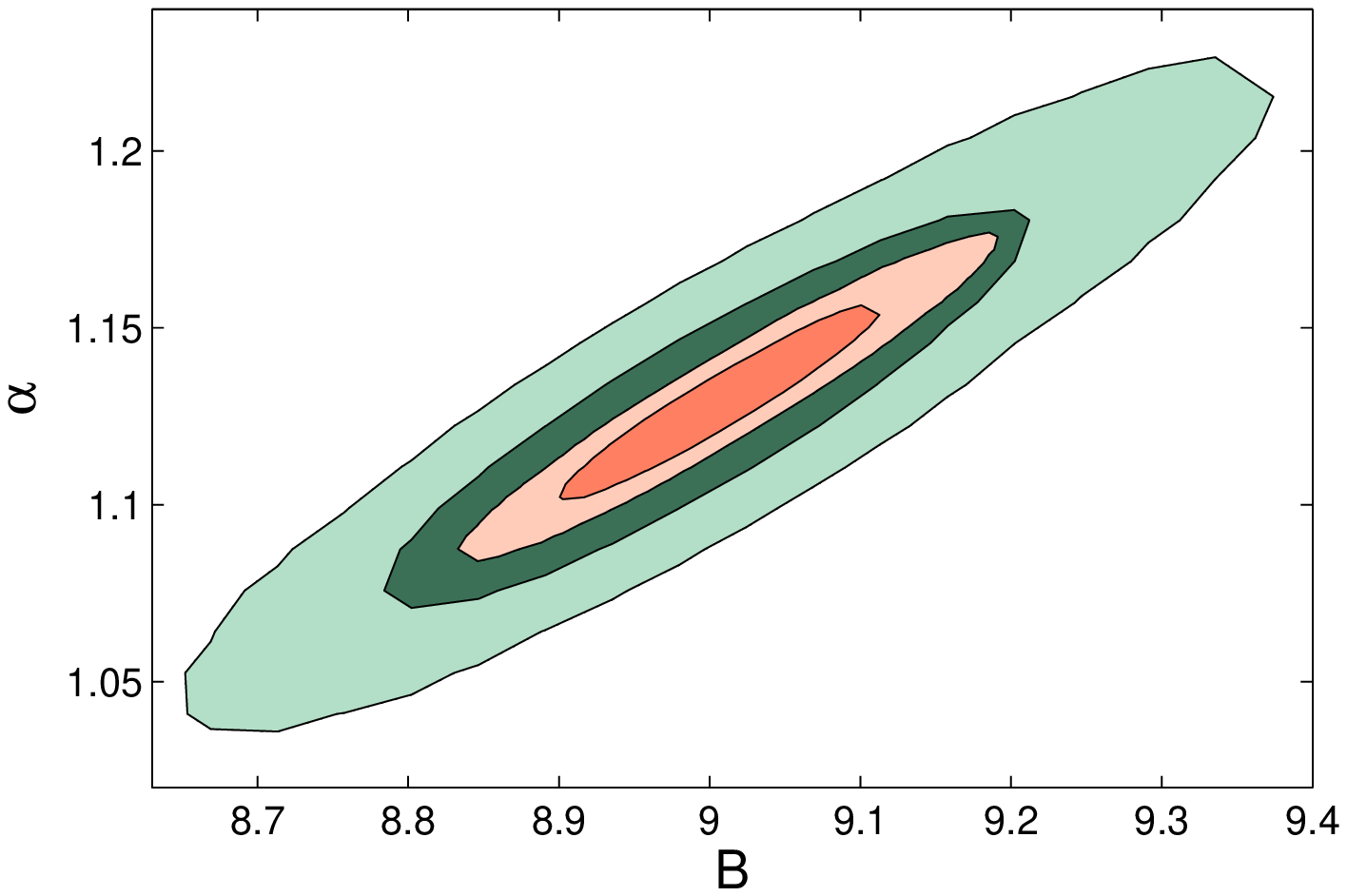}
  \label{fig:confd}
}
\subfigure[]{
  \includegraphics[width=8cm]{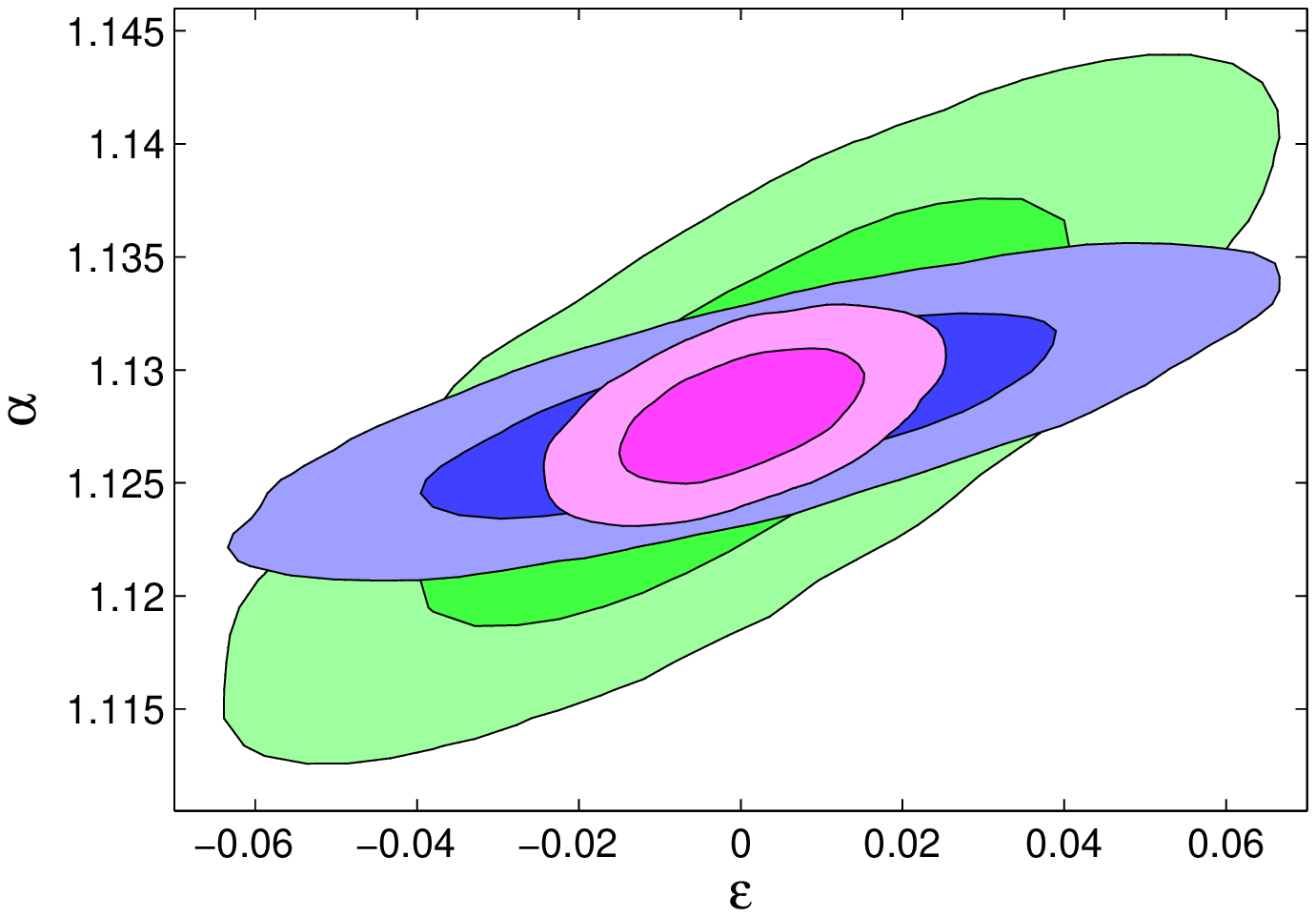}
}
\vspace*{-6mm}
\caption[]{1$\sigma$ \& 2-$\sigma$ marginalized constraints from 21 cm galaxy surveys -- 
(Left) The green regions indicate parameter constraints obtained from 
observed ADBS galaxies (in which we fix $\epsilon$=0), while the orange regions represents 
constraints (marginalized over $\epsilon$) from a simulated galaxy 
catalog expected from the $5\times$eVLA survey. (Right) The green/blue regions correspond to the expected constraints from
simulated galaxies in the SKA deep/wide surveys respectively. Note, how 
the degeneracy is broken (magenta) by combining a deep and wide survey.
}
\end{figure*}

Properties of the Galaxies: The luminosity, $L$ and area, $A$ of disk galaxies are seen to be related
almost linearly \cite{2003ApJ...585..256R, 2011arXiv1103.2763C},
\begin{equation}
 L = 10^{B} \left( A/A_0 \right) ^{\alpha} {\rm Jy \: km \: s^{-1} \: Mpc^2}
\label{eqn:linear}
\end{equation}
where we choose $A_0$ = 1 Mpc$^2$. Fig. \ref{fig:confd} shows the 
constraints on the parameters $B$ and $\alpha$ with the existing 
data of the ADBS galaxies.

We use the Schechter mass function from \citet{2005MNRAS.359L..30Z} to
model the HI mass distribution at low redshifts,
\begin{equation}
\frac{\rm dN}{\rm dVdM_{HI}} = \frac{\theta^{*}}{M^*_{\rm HI}} {\left(
\frac{M_{\rm HI}}{M^{*}_{\rm HI}} \right)}^{\beta} \exp \left(-
\frac{M_{\rm HI}}{M^{*}_{\rm HI}} \right)
\label{eqn:schechter}
\end{equation}
with $\beta=-1.37$ and $\log \left( M^{*}_{\rm HI} / {\rm M}_{\odot} \right) = 9.80$.
We hold $M^{*}_{\rm HI}$ constant as it is likely to be regulated by the
mass scale beyond which AGN feedback dominates supernova feedback \cite{2005MNRAS.364.1337S}.
However, as HI mass is directly related to the star formation rate, we choose a
redshift dependent normalization for the HI mass function; $\theta^{*}(z) = 6
\times {\rm 10^{-3}}(1+7.6z)/(1+(z/3.3)^{5.3})$, such that it matches the shape of SFR($z$)
from \citet{2006ApJ...651..142H}.
Any uncertainty in modeling the mass function here, only affects 
the number of galaxies yielded by the simulated survey and has no effect
on the actual Tolman test.

Following \citet{2000tra..book.....R},
 we relate the HI masses to the flux $f$ as, $M_{\rm HI}/M_{\odot}=2.356 \times
{\rm 10^5} \: f d_L^2 (1+z)^{-1}$ \footnote{The extra factor of 
$(1+z)$ comes up in radio astronomy because the flux
density (which is measured per Hz) is integrated over the velocity
(in km s$^{-1}$) as opposed to the usual frequency (in Hz).}, 
where $f$ is in Jy km s$^{-1}$ and $d_L$ is in Mpc.

Corresponding to the the flux cut of a given survey, we calculate 
the limiting luminosities $L(z)$ and corresponding masses $M_{\rm HI}^{\rm (lim)}(z)$.
The flux from the galaxy is assumed to be spread out, due to rotation, over
a velocity range of $\Delta V=$ 200 km s$^{-1}$. If one assumes the weak scaling
($\Delta V \propto M_{\rm HI}^{0.3}$) from \citet{1993ApJ...417..494B} 
and inclination effects, one would expect $\sim30\%$ more
sources \citep{2011arXiv1103.3944D}.
In addition to the flux cut-off we also impose a size cut-off according to the beam size of the survey.
The number of galaxies above the limiting mass of the survey with a coverage area $\Delta \Omega$ 
in a redshift bin $\Delta z$ is given by,
\begin{equation}
\Delta \Omega \Delta z \frac{\rm dV}{{\rm d}z {\rm d}\Omega} \int_{M_{\rm HI}^{\rm
(lim)}(z)}^{\infty} \frac{\rm dN}{{\rm dV} {\rm d}M_{\rm HI}} {\rm d}M_{\rm HI}
\label{eqn:Ngal}
\end{equation}
where ${\rm V}$ is the comoving volume upto a redshift $z$. This gives the 
luminosity and redshift distribution of galaxies and is used to create 
the simulated catalogs of galaxy data containing $z$, $L$, $S$ and $\sigma(S)$, where
 $\sigma(S)$ obtains contribution mostly from the intrinsic scatter 
in the data as calibrated from the ADBS galaxies.

Surveys at 21 cm: The properties of the simulated surveys are presented in Fig. \ref{tab:both}.
``ADBS + VLA'' represents the constraints on $B$ and $\alpha$ obtained using
the real data \citep{2003ApJ...585..256R}. ``5$\times$eVLA'' represents
results from a simulated catalog, assuming 80 hrs of on-source time using
a hypothetical telescope with 5 times the sensitivity of the present Extended VLA (eVLA).
This may soon become a reality with upgrades at the eVLA or upcoming pathfinder
missions of the SKA.
``SKA {\it deep}'' represents a single snapshot (2 minutes) with the fully
operational SKA. ``SKA {\it wide}'' represents the same amount of SKA time, but
spent on 100 different fields instead of 1. While the SKA {\it deep}
is good for constraining any redshift evolution in the CDDR,
the SKA {\it wide} is good for constraining the actual size luminosity
relation of the galaxies.

{\it Results}.---We modify the Etherington relation to test for any violation by using a simple parametrized form,
$d_L/d_A=(1+z)^{2 +\epsilon}$. Using the definitions $d_A^2(z) = A/S$  and 
$d_L^2(z) = (1+z) L / (4 \pi f)$ \footnotemark[\value{footnote}], we 
relate the observed angular area ($S$) to observed flux ($f$) as follows,
\begin{equation}
S(z,f) = \left(\frac{4\pi f (1+z)^{3+2 \epsilon}}{10^B
d_A(z)^{2(\alpha-1)}}\right)^{\frac{1}{\alpha}}
\label{eqn:size-flux}
\end{equation}
\normalsize

The almost linear relation ($\alpha\simeq1$) between HI masses and surface area of disk galaxies
implies that the residual cosmological dependence (through $d_A(z)$) in the
above equation is only expected to be weak. For the cosmology we assume the standard $\Lambda$CDM 
model \footnote{The constraints on $\epsilon$ are only weakly affected by a wrong model of 
cosmology; eg. in the SKA {\it deep}+{\it wide} survey, an analysis with a  1$\sigma$ error in a model with
 fixed $w$ gives a bias of only 0.003 in $\epsilon$, while for the corresponding error in $\Omega_k$ we
 see an even smaller bias of $<$ 0.0005 in $\epsilon$. Marginalizing over the parameters $\Omega_k$ and $w$, 
with priors from the WMAP7 results, in an open $w$CDM model does not change our results appreciably.}
 keeping priors (from WMAP7 results \cite{2011ApJS..192...18K}) on $h$ and
$\Omega_m$. We perform a MCMC likelihood analysis of the simulated data with the parameters 
-- $B$, $\alpha$, $\epsilon$, $h$ and $\Omega_m$. The likelihoods
 from the data are computed as $\exp(-\chi^2/2)$, where 
$\chi^2=\displaystyle\sum_{i}^{} \left(\log(S(z_i,f_i)) - \log(S_i) \right)/
\sigma_i^2$.

We first analyze the ADBS data keeping the parameter $\epsilon$ to be fixed,
 since the existing data is at very low redshifts ($z<0.025$) which is clearly 
insufficient to test for violations of the CDDR. We find 
the best fit parameters values as $B$=9.003 and $\alpha$=1.128 (see Fig. \ref{fig:confd}).
We use these values for $B$ and $\alpha$ along with the fiducial value of $\epsilon$=0 
to create the mock catalog following the procedure as explained previously.

We then analyze the simulated catalogs for some of the upcoming 21 cm observations like 
 5$\times$eVLA and the SKA. All of these surveys would throw up large number of galaxies
 in detection, which means that we may simply self-calibrate the scaling relation 
 of 21 cm luminosity with area (see eq. (\ref{eqn:linear})). We discuss the prospects of 
the surveys to probe the CDDR by obtaining marginalized constraints
 on $\epsilon$, the parameter for violation of this relation.

The marginalized constrains on $\epsilon$ from 5$\times$eVLA would be comparable 
to some of the present constraints \cite{2009JCAP...06..012A}, 
with $\Delta \epsilon$=0.095. These constraints can be significantly 
improved by the next generation telescopes like the SKA. The SKA deep survey with a very good flux 
sensitivity is expected to detect galaxies upto a high redshift of $z$ $\sim 1$. The 
100 sq. deg. wide SKA survey would have many more objects, however, because of its 
higher flux cut, it would mostly detect low mass galaxies at lower redshifts, $z$
 $\sim 0.4$ (see Fig. \ref{tab:both}). Both the deep and wide surveys from the SKA
 would give $\Delta \epsilon$=0.026; however, because these surveys probe the relation 
in eq. (\ref{eqn:linear}) at very different range of masses they place constraints along 
different directions. This fact can be used to a great advantage by using the deep and 
the wide surveys in conjunction; as shown in Fig. \ref{fig:confd}, this is excellent at breaking 
the degeneracy between $\epsilon$ and $\alpha$. This {\it wedding cake} \cite{2010PhRvD..82d1301K}
 survey strategy would strongly constrain $\epsilon$ to 0.01 
while also constraining the relation in eq. (\ref{eqn:linear}) to a high precision with 
$\Delta B$=0.007 and $\Delta \alpha$=0.002. 

{\it Discussions}.---The planned SKA will allow the detection of 21 cm emission from
distant ($z\sim1$) galaxies with small observing times.
Apart from being useful in performing the Tolman test, the same
data will be useful for cosmology and galaxy evolution studies. 
The deepest observations may allow one to reach $z\sim6$ and start
probing the HI content of the first galaxies, which would have
played a vital role in the epoch of reionization.
 
Self regulation of mechanical feedback from supernovae and gravitational
instability of the star forming disks keeps the surface density nearly
constant, irrespective of the global star formation rates.
The almost constant observed surface density of HI in disk galaxies makes them
ideal sources for the Tolman test -- Firstly, by using the same objects to directly probe the
ratio $d_L/d_A$ at various redshifts, the results are not 
affected by many possible astrophysical model uncertainties.
Secondly, our test also promises to be immune to bias from effects
of gravitational lensing as it preserves the surface 
brightness of such sources. Lastly, our proposed test does not
fix the scaling relation between luminosity and the area of HI regions
in galaxies; the upcoming surveys would simply detect enough number
of these galaxies to permit a self-calibration of this
relation.

We have shown that observations at the 21 cm wavelength opens up
the exciting possibility of testing the CDDR in
a completely independent manner. Probing this relation with our proposed method,
 offers several advantages over all of the existing tests. At radio 
frequencies the universe is essentially transparent. Thus one can 
cleanly distinguish between violations of this relation 
arising from astrophysical effects and those that may arise from new physics.
Our proposed test of the CDDR may soon enable 
us to probe the possibility of photons coupling to the dark sector.

{\it Acknowledgements}.---The authors would like to thank Satyabrata Sahu for detailed discussions,
suggestions and a careful reading of the manuscript. Biman Nath, Subhabrata Majumdar, Surhud More,
 Bruce Bassett, Martin Kunz are thanked for giving valuable comments on this manuscript.


\end{document}